\documentclass[11pt]{article}
\usepackage{amsmath,amsfonts, braket}
\usepackage{tikz}
\usetikzlibrary{trees,er,snakes,shapes,mindmap}
\usepackage{sectsty}
\sectionfont{\fontfamily{put}\fontseries{b}\fontshape{n}\fontsize{12pt}{14pt}\selectfont}

\def \beq  {\begin{equation}}
\def \eeq  {\end{equation}}
\def \beqar {\begin{eqnarray}}
\def \eeqar {\end{eqnarray}}
\allowdisplaybreaks
\def\sqr#1#2{{\vcenter{\vbox{\hrule height.#2pt
\hbox{\vrule width.#2pt height#1pt \kern#1pt
\vrule width.#2pt}\hrule height.#2pt}}}}

\def\vx {{\vec x}}

\def\vf {{\varphi}}

\def\Tr {{\rm Tr}}

\def\bD {\bar{D}}
\def\bA {\bar{A}}

\def\vx {{\vec x}}

\def\del {\partial}
\def\bdel{\bar{\partial}}

\def\l {\lambda}

\def\bz {{\bar{z}}}

\def\A {{\cal A}}

\def\I{{\cal I}}

\def\M{{\cal M}}

\def\vf {{\varphi}}

\def\half{\textstyle{1\over 2}}

\begin{document}
\fontfamily{put}\fontsize{11pt}{16pt}\selectfont
\def \CMP {{Commun. Math. Phys.}}
\def \PRL {{Phys. Rev. Lett.}}
\def \PL {{Phys. Lett.}}
\def \NPBProc {{Nucl. Phys. B (Proc. Suppl.)}}
\def \NP {{Nucl. Phys.}}
\def \RMP {{Rev. Mod. Phys.}}
\def \JGP {{J. Geom. Phys.}}
\def \CQG {{Class. Quant. Grav.}}
\def \MPL {{Mod. Phys. Lett.}}
\def \IJMP {{ Int. J. Mod. Phys.}}
\def \JHEP {{JHEP}}
\def \PR {{Phys. Rev.}}
\def \JMP {{J. Math. Phys.}}
\def \GRG{{Gen. Rel. Grav.}}
\begin{titlepage}
\null\vspace{-62pt} \pagestyle{empty}
\begin{center}
\rightline{October 2019}
\vspace{1truein} {\Large\bfseries
Sasakians and the Geometry of a Mass Term}\\
\vskip .15in
{\Large\bfseries ~}\\
\vskip .1in
{\Large\bfseries ~}\\
 {\large\sc V.P. Nair}\\
\vskip .2in
{\itshape Physics Department,
City College of the CUNY\\
New York, NY 10031}\\
 \vskip .1in
\begin{tabular}{r l}
E-mail:&\!\!\!{\fontfamily{cmtt}\fontsize{11pt}{15pt}\selectfont vpnair@ccny.cuny.edu}\\
\end{tabular}
\vskip 1in
\centerline{\large\bf Abstract}
\end{center}
A gauge-invariant mass term for nonabelian
gauge fields in two dimensions can be expressed as the Wess-Zumino-Witten (WZW) action.
Hard thermal loops in the gauge theory in four dimensions at finite
temperatures generate a screening mass for some components of the gauge
field. This can be expressed in terms of the WZW action
using the bundle of complex structures (for Euclidean signature)
or the bundle of lightcones over Minkowski space.
We show that a dynamically generated mass term
in three dimensions can be put within the same general framework using
using the bundle of Sasakian structures.
\vskip .2in
\centerline{\it Dedicated to Roman Jackiw on his 80th Birthday}
\noindent To appear in "Roman Jackiw: 80th Birthday Festschrift", edited by A. Niemi, T. Tomboulis, and K. K. Phua (World Scientific, 2020)
\end{titlepage}
\pagestyle{plain} \setcounter{page}{2}
\section{Introduction}
The early 1980s were a time of growing appreciation of the role of topology
in quantum field theory, especially for gauge theories. Anomalies and Chern-Simons terms
were very much in the air, so it was impossible for any graduate student 
to be unaware of the seminal contributions of Roman Jackiw. My own collaboration with 
Roman began somewhat later, during his sabbatical visit to Columbia University in 1990. At that time Roman was very much interested
in solitons in Chern-Simons theories coupled to matter fields, both relativistic and nonrelativistic
\cite{JP-soliton}, but we did talk about anomalies and anyons and representations of the Poincar\'e group
in 2+1 dimensions.
After he went back to MIT, we corresponded about anyons, and this evolved into our paper on the relativistic wave equation for anyons \cite{JN1}.

Throughout the 1990s and early 2000s, we continued to
collaborate on a number of projects of common interest, from finite temperature field
theories,
nonabelian Clebsch parametrization \cite{JN-clebsch}, a group theory based formulation of
nonabelian magnetohydrodynamics \cite{hydro}, etc.
Particularly gratifying was my work with Efraty on an effective action for hard thermal
loops \cite{nair-htl}, and the subsequent work with Roman on developing it into
a nonabelian version of the Kubo formula \cite{JN-htl2}, combining two of his
favorite topics: field theory at finite temperature and Chern-Simons theory.
Chromomagnetic screening masses and gap equations in 2+1 dimensional gauge theories
was another topic on which I had many discussions with Roman and So-Young,
although we never published any joint work on this \cite{JP-magmass}.
Looking back, it is striking to me that we had overlap of interest on so many different topics.
Yet, on second thought, it is perhaps not so remarkable since Roman has been 
a continuing influence on the development of field theory from the mid-1960s to the present, and
hence any one interested in field theory would be bound to have many points of
overlap with his work.

To a man who has devoted decades to physics, appreciation must be shown in kind,
not just in anecdotes and reminiscences alone.
So, for my contribution to this Festschrift,
 I have decided to write on a novel aspect of something we have both worked on.
I shall discuss dynamically generated mass terms in gauge theories, 
which brings together Chern-Simons actions and their eikonals, the Wess-Zumino-Witten actions,
Dirac determinants,
chromomagnetic screening effects and many facets of geometry and topology,
which are all topics of interest to Roman.
The key point is that while K\"ahler structures play an important role
for physics in even
dimensions, Sasakian structures should do so in odd dimensions. 
A mass term which can be dynamically generated in nonabelian gauge theories
in odd dimensions,
as I argue below, exemplifies this.

\section{Masses for gauge theories in two and four dimensions}

The prototypical example of a gauge-invariant mass term
is given by the Wess-Zumino-Witten (WZW) action in two dimensions
\cite{witten}, or, equivalently, by the logarithm of the Dirac determinant \cite{PW}.
This is the nonabelian generalization of Schwinger's result for the Abelian case
\cite{schwinger}.
Specifically, this mass term takes the form
\beqar
\Gamma &=& - [\Tr \log \left( - \bD D \right) - \Tr \log (-\bdel \del )] = - A_R \,S_{\rm WZW}(H)\nonumber\\
S_{\rm WZW} (H) &=& {1\over 2\pi} \int_\M d^2z \, \Tr (\del H \bdel H^{-1} )\nonumber\\
&&+ {i \over 12\pi} \int_{\M^3}\Tr (H^{-1} dH \wedge H^{-1} dH \wedge H^{-1} dH )
\label{mass1}
\eeqar
where we use complex coordinates in two dimensions, 
$z, \, \bz = x_1 \mp i x_2$ and 
\beqar
D&=& \del + A = {\half} (\del_1 + i \del_2) +   {\half} (A_1 + i A_2) 
= {\half} (\del_1 + i \del_2)  + (-it_a ) {\half} (A_1^a + i A_2^a )
\nonumber\\
\bD &=& \bdel + \bA = {\half} (\del_1 - i \del_2) +   {\half} (A_1 - i A_2) 
= {\half} (\del_1 - i \del_2)  + (-it_a ) {\half} (A_1^a - i A_2^a )
\label{mass1b}
\eeqar
One can parametrize the gauge field in terms of a complex matrix $M$ as
$A = - \del M M^{-1}$, $\bA = M^{\dagger -1} \bdel M^\dagger$, which yields the second expression
in (\ref{mass1}) in terms of the WZW action with $H = M^\dagger M$.
In (\ref{mass1b}),
$\{ t_a\}$ are a set of hermitian matrices forming a basis for the Lie algebra of the gauge group,
$A_R = 1$ if the fields are in the fundamental representation (F),
otherwise, for representation $R$,  it is defined by
$\Tr (t_a t_b )_R = A_R \, \Tr (t_a t_b )_F$. The integration is over the two-dimensional space 
$\M$ of
interest; in the second term of $S_{\rm WZW}$ we extend the fields to a three-manifold
$\M^3$ whose boundary is $\M$, as usual. 

The WZW action we have written can also be expressed directly in terms of
the gauge potentials, which is useful for explicit computations in a gauge theory.
It reads
\beqar
S_{\rm WZW} (H) &=& {1\over \pi} \left[\int d^2z\, \Tr (A \bA ) - \pi \I (A) -\pi {\bar \I}(\bA)\right]
\nonumber\\
\I(A) &=& \sum_2^\infty {(-1)^n \over n\, \pi^n} 
\int d^2z_1  \cdots d^2z_n {\Tr \bigl( A(z_1, \bz_1) A(z_2, \bz_2) \cdots A(z_n, \bz_n) \bigr)
\over (\bz_1 - \bz_2) (\bz_2 - \bz_3) \cdots (\bz_n - \bz_1)}\label{mass1a}
\eeqar
and ${\bar \I}(\bA)$ is similar with $A \rightarrow \bA$, $\bz_1- \bz_2 \rightarrow z_1 - z_2$,
etc. for the terms in the denominator of the expression in (\ref{mass1a}).

The fact that we have a complex structure for ${\mathbb R}^2$ is important in constructing this
mass term. The WZW action, and hence the mass term, can be written for any 
Riemann surface, viewed as a complex manifold, by
a simple generalization as
\beqar
S_{\rm WZW} (H) &=& {1\over 8\pi} \int_\M d^2x \,\sqrt{g}\, g^{ab}\, \Tr (\del_a H \del_b H^{-1} )
\nonumber\\
&& \hskip .2in + {i \over 12\pi} \int_{\M^3} \Tr (H^{-1} dH \wedge H^{-1} dH \wedge H^{-1} dH )
\label{mass2}
\eeqar
where $g_{ab}$ is the metric tensor for the two-manifold.

Consider now the extension of this to four dimensions. A mass term, similar to (\ref{mass1}),
can be written down if we can identify two complex coordinates out of the four real coordinates of
${\mathbb R}^4$. But there are many choices for a complex (or even a K\"ahler) structure.
We can understand the inequivalent choices as follows. If we choose one set of complex combinations, say
$\omega_1 = x_0 +i x_3$, $\omega_2 = x_2 -i x_1$, then a $U(2)$ transformation
of $(\omega_1, \omega_2)$ does not change the complex structure.
In particular, a holomorphic function of $\omega = (\omega_1, \omega_2)$ remains holomorphic
under the $U(2)$ transformation.
However, we can do a rotation of all four coordinates, as $x_\mu \rightarrow x_\mu' = R_\mu^{~\nu}
x_\nu$, where $R_\mu^{~\nu}$ is a rotation matrix,
and this does lead to a different structure given by
$x'_0 +i x'_3,\, x'_2 -i x'_1$.
So the inequivalent ways of combining $(x_0, x_1, x_2, x_3)$ into complex combinations
are parametrized by $SO(4)/ U(2) \sim S^2$.
More explicitly, introduce a two-spinor $(\pi_1, \pi_2)$
with the identification $\pi = (\pi_1, \pi_2) \sim \lambda (\pi_1, \pi_2)$, $\lambda \in {\mathbb C} - \{ 0\}$.
The $\pi$'s parametrize ${\mathbb{CP}}^1 \sim S^2$.
The complex combinations can be taken as
\beq
\left( \begin{matrix} \omega_1 \\ \omega_2\\ \end{matrix} \right)
= (x_0 -i \sigma_i x_i ) \pi
= \left[ \begin{matrix} x_0 -i x_3& -x_2-i x_1\\ x_2 -i x_1& x_0 +i x_3\\ \end{matrix}
\right] \left( \begin{matrix} \pi_1 \\ \pi_2\\ \end{matrix} \right)
\label{mass3}
\eeq
where $\sigma_i$ are the Pauli matrices. We can also define a real unit vector
$Q_i = {\bar \pi} \sigma_i \pi /({\bar\pi}\pi)$.
We then find
\beq
({\bar \pi} \omega, {\bar \omega} \pi ) = (x_0 -i {\vec Q}\cdot \vx , ~x_0 +i {\vec Q}\cdot \vx )
\equiv (z, \bz )
\label{mass4}
\eeq
These constitute two of the complex coordinates. The remaining two transverse coordinates are
given by $\vx \times {\vec Q}$. The unit vector $Q_i$ gives an alternate way to parametrize
$S^2$. For any fixed choice of ${\vec Q}$, we do lose rotational invariance
but we can construct an invariant mass term as \cite{nair-htl}
\beq
\Gamma = - k \int d\mu_{S^2}\, d^2x^T~ S_{\rm WZW} (H)
\label{mass5}
\eeq
where $H = M^\dagger M$ and
\beqar
A&=& {\half} ( A_0 +i {\vec Q}\cdot {\vec A} ) = - \half
 ( \del_0 +i {\vec Q}\cdot {\nabla} ) M\, M^{-1}\nonumber\\
 \bA&=& {\half} ( A_0 -i {\vec Q}\cdot {\vec A} ) =  \half
 M^{\dagger -1} ( \del_0 -i {\vec Q}\cdot {\nabla} ) M^\dagger
 \label{mass6}
 \eeqar
 The integrations over all orientations of ${\vec Q}$, signified in (\ref{mass6}) by 
 $d\mu_{S^2}$, and over the transverse coordinates $x^T$, will make this mass
 term rotationally invariant. The integration over $z, \bz$ is part of $S_{\rm WZW}(H)$, 
 so the final result in (\ref{mass5})
 will have integration over all four coordinates with the measure $d^4x$.
This mass term is also obviously gauge-invariant, in the same way as in
 two dimensions.
 
A number of comments are in order at this point. First of all,
the key idea here is to use a pair of complex coordinates or more 
generally a two-dimensional complex subspace to construct the WZW action.
This necessarily entails a lack of rotational symmetry. Symmetry is restored by integrating over all
possible choices of complex coordinates. In other words, we may think of the total space
of interest
as the bundle of complex structures on ${\mathbb R}^4$. This is basically the way twistor space
is defined \cite{twistor}.
 In the present case, where the base space is flat, the bundle is trivial. One could do 
an analogous construction for other spaces, a notable example being $S^4$. For this latter case, 
we cannot have a global complex structure, so combinations of coordinates into complex ones correspond to local complex structures and one is considering the bundle of local 
complex structures
over $S^4$. The bundle space is then ${\mathbb{CP}}^3$, with $S^2$ as the fiber and 
$S^4$ as the base, and the bundle is topologically nontrivial.
 In any case, the key point here is to consider the bundle of
local complex structures, trivial or nontrivial.

Secondly, while the twistor space genesis of (\ref{mass5}) may be mathematically gratifying, 
one may ask whether this
mass term has anything to do with physics. The remarkable fact is that it does.
Of course, its use in physics needs a continuation to Minkowski signature, not
the Euclidean one we have used so far. This continuation can be done by
the rules
\beqar
A&\rightarrow& {\half} ( A_0 +  {\vec Q}\cdot {\vec A} ) = - \half
 ( \del_0 +  {\vec Q}\cdot {\nabla} ) M\, M^{-1}\nonumber\\
 \bA&\rightarrow& {\half} ( A_0 - {\vec Q}\cdot {\vec A} ) =  \half
 M^{\dagger -1} ( \del_0 - {\vec Q}\cdot {\nabla} ) M^\dagger
 \label{mass7}
 \eeqar
 If we consider a physical system described by a nonabelian gauge theory such as
quantum chromodynamics (QCD), then, at finite temperature where we get a plasma of 
gluons (and quarks if they are included), the terms in the standard perturbative expansion have
infrared divergences. There is an infinite sequence of terms, which are the leading infrared divergent terms, known as hard thermal loops (HTL) \cite{htl}. These are special to the case of
nonzero temperature and are in addition to the usual divergences (both in the ultraviolet and infrared) in the theory at zero temperature.
These HTL terms have to be summed up and included at the lowest order
to reformulate perturbation theory without infrared divergences. (This has to be done in a 
self-consistent way, the technology for this is well understood.)
The summation of the HTL terms is a screening effect
 for the electric-type forces corresponding to the nonabelian
 gauge field. In fact it is the nonabelian generalization of the Debye screening effect well known
 for the Abelian plasma (and electrolytes).
 The sum of the HTL terms
 can be interpreted as a mass term, primarily
 for the $A_0$-component of the gauge field,
 with some contributions to the other components as well to satisfy the Gauss law.
 The HTL contributions can be calculated in the field theory at finite temperature
 and the
result of the detailed calculations at finite temperature is exactly
 the mass term (\ref{mass5}), with the continuation
 in (\ref{mass7}), and with $k = (N + {\half } N_F ) T^2/6$ \cite{nair-htl}. This value of $k$
 is for the case of an 
 $SU(N)$ gauge theory, with $N_F$ massless fermion flavors and
 $T$ denotes the temperature of the plasma.
 Thus, quite remarkably, what was defined purely as a mathematical generalization
 is indeed realized in explicit calculations in a very physical context, namely, the quark-gluon
 plasma.
 
A similar screening effect also occurs for a degenerate gas of quarks with a nonzero
total baryon number, such as can occur deep inside a neutron
star. This is the nonabelian generalization of the well-known
Thomas-Fermi screening
effect for electron gases. The mass term describing this
 is again of the same form, with
$k= \mu_q^2 /4\pi^2$, where $\mu_q$ is the chemical potential for the 
quark number ($={1\over 3}$ of the baryon number) \cite{alexanian-nair}.

Finally, we may raise the question of Lorentz invariance. The Minkowski continuation
of the mass term as written
in (\ref{mass5}) is not Lorentz-invariant. Physically, this is indeed as it should be, since thermal 
equilibrium and the specification of the temperature are obtained in the rest frame of 
the plasma without any overall drift velocity. For a Lorentz-invariant result, we need one more
parameter, the overall drift velocity of the plasma, whose Lorentz transformation will lead to an
invariant result. The relevant form of the mass term (i.e., the generalization for the moving plasma)
was worked out many years ago
and takes the form \cite{nair-moving}
\beqar
\Gamma &=& - k \int d\mu\, d^2x^T\, S_{WZW}(H)\nonumber\\
d\mu&=& 2 i  {{\pi\cdot d\pi}\,  {{\bar \pi} \cdot {d {\bar \pi} }\, {\xi\cdot d\xi} \,{{\bar \xi} }\, \cdot d {\bar \xi}}
\over (\pi \cdot \xi )^2 ( {\bar \pi} \cdot {\bar \xi})^2}
~\delta[ \xi (e\cdot p) {\bar\pi} ]~ \delta [ \pi (e \cdot p) {\bar \xi} ]
\label{mass8}
\eeqar
We have introduced two sets of two-component $SL(2, {\mathbb C})$ spinors, $\pi^A$, $\xi^A$, $A= 1,2$,
with $\pi^{\dot A} = \overline{\pi^A}$ and $\xi^{\dot A} = \overline{\xi^A}$ as their complex conjugates.
The components of the gauge fields used to define $M$ and $M^\dagger$, and hence $H$,
are given by
\beq
A_\pi = {\half} \pi (e\cdot A) {\bar \pi} , \hskip .2in
A_\xi = {\half} \xi (e \cdot A) {\bar \xi}
\label{mass9}
\eeq
Further $e^\mu = ( {\mathbf 1}, \sigma^i )$, and $p_\mu$ in (\ref{mass8})
denotes the drift $4$-velocity of the 
plasma. The derivatives are defined in a way similar to the $A$'s given above, namely by
(\ref{mass9}) with $A \rightarrow \del$, with corresponding expressions for the coordinates. Notice the presence of the $\delta$-functions in $d\mu$.
Upon integration, they enforce a relation between the two spinors in a way which depends on
$p^\mu$. In the rest frame of the plasma, with $p^\mu = (1, 0, 0, 0)$, expression (\ref{mass8})
reproduces the previous result (\ref{mass5}).

Another feature worthy of remark is that the combinations of the gauge potentials
in (\ref{mass7}), as well as derivatives, can be written
as $n \cdot A $, $n\cdot \del$ and $n'\cdot A$, $n'\cdot\del$, where
$n^\mu = (1, i Q_i )$ $n'^\mu = (1, -i Q_i ) = {\bar n}^\mu$ in Euclidean space.
These are complex null vectors. Upon continuation to Minkowski space, we get
$n^\mu = (1,  Q_i )$ $n'^\mu = (1, - Q_i )$, which are real null vectors.
Thus the bundle space we are considering is the bundle of lightcones over 
Minkowski space \cite{twistor}.

Turning now to three-dimensional space, obviously we cannot combine coordinates
 pairwise into complex combinations, so an immediate generalization seems difficult.
 However, there is a mathematical structure known as the Sasakian
 which can exist on certain odd-dimensional manifolds and
 which has been suggested as the closest we can get to a K\"ahler structure.
 We can try to utilize this to construct a mass term.
 In the next section, we give a general discussion of Sasakians for
 $S^3$ and ${\mathbb R}^3$ and write down this mass term.
 The final result agrees with what was suggested as the magnetic screening mass
 for the gluon plasma many years ago, although the Sasakian connection was not
 apparent at that time.

\section{$S^3$ and ${\mathbb R}^3$ as Sasakian manifolds and a 3d mass term}

We begin by briefly recalling the definition of a Sasakian manifold \cite{sasakians}.
Let $\M$ be an odd-dimensional  Riemannian manifold with the metric
$ds^2_\M$. The Riemannian cone for $\M$ is
$\M \times {\mathbb R}_+$
with the cone metric
\beq
ds^2 = dr^2 + r^2 \, ds^2_\M
\label{mass10}
\eeq
where $r \in {\mathbb R}_+$ is the additional coordinate along
the ${\mathbb R}_+$ direction. The manifold $\M$ is said to be a contact manifold
if there is a one-form ${\hat \Theta}$ on $\M$ such that the two-form
\beq
\Omega = r^2 d {\hat \Theta} + 2 r dr\, {\hat \Theta}
\label{mass11}
\eeq
is symplectic. The manifold $\M$ endowed with ${\hat \Theta}$ is Sasakian
if the two-form $\Omega$ and the metric $ds^2$ on the cone, i.e., (\ref{mass10}), are K\"ahler.
Since $\M$ is a transverse cross-section of the cone, it inherits many properties
from the K\"ahler structure of the cone. In fact, it is generally considered that 
the Sasakian structure is the closest one can get to K\"ahler-type properties for an odd-dimensional space.

We can apply this specifically to $S^3$ by considering its embedding in ${\mathbb R}^4$
and taking $r$ as the radial coordinate. Removing the origin, ${\mathbb R}^4 - \{ 0\}$ has the cone structure, with the metric on the cone being the flat Euclidean metric
$ds^2 = dx_0^2 + dx_1^2 + dx_2^2 + dx_3^2$. To identify $S^3$ as a Sasakian space, we need to write this
metric as a K\"ahler metric. As discussed in the last section, there are an infinity of
 inequivalent ways of doing this, the possible complex combinations being parametrized
 by $\pi$ which form the homogeneous coordinates for ${\mathbb{CP}}^1$.
 Using the freedom of scaling, $\pi \sim \l \pi$, $\l \in {\mathbb C}- \{0\}$, we can bring it to the form
 \beqar
 \left( \begin{matrix} \pi_1 \\ \pi_2 \\ \end{matrix} \right)
 &=&  \left( \begin{matrix} -e^{-i {\vf\over 2}}\cos{\theta\over 2}  \\ -e^{i {\vf\over 2}}\sin{\theta\over 2}  \\ \end{matrix} \right)
 \nonumber\\
 &=&\left[ \begin{matrix} - e^{-i {\vf\over 2}} \sin{\theta\over 2}&~-e^{-i {\vf\over 2}}\cos{\theta\over 2}\\
e^{i {\vf\over 2}} \cos{\theta\over 2}& ~- e^{i {\vf\over 2}}\sin{\theta\over 2}\\ \end{matrix}
 \right]\, \left( \begin{matrix} 0\\ 1\\ \end{matrix} \right) \equiv g \left( \begin{matrix} 0\\ 1\\ \end{matrix} \right)
 \label{mass12}
 \eeqar
The complex combinations for ${\mathbb R}^4$ can be taken as
 in (\ref{mass3}). But as mentioned earlier, we are free to do a $U(2)$ rotation
 of the complex combinations without changing the complex structure. For our purpose here,
 it is useful to do this using $g^\dagger$, thus defining
 \beqar
 \omega &=& g^\dagger ( x_0 - i {\vec \sigma} \cdot \vx) \pi 
 = g^\dagger ( x_0 - i {\vec \sigma} \cdot \vx) g \,\left( \begin{matrix} 0\\ 1\\ \end{matrix} \right)
 \nonumber\\
&=&(x_0 -i \sigma_k R_{ki} x_i ) \left( \begin{matrix} 0\\ 1\\ \end{matrix} \right)
\label{mass13}
\eeqar
For the choice of $\pi$ in (\ref{mass12}), the components of the orthogonal matrix $R_{ki}$ are given by
\beq
R_{ki} = \left[ \begin{matrix}
-\cos\theta\, \cos\vf& - \cos\theta\,\sin\vf& \sin\theta\\
-\sin\vf& \cos\vf& 0\\
-\sin\theta\, \cos\vf&-\sin\theta\, \sin\vf &-\cos\theta\\
\end{matrix} \right]
\label{mass14}
\eeq
In terms of the complex coordinates $\omega$, the K\"ahler forms and metric can be taken
as
\beqar
\Omega&=& i \left(d{\bar \omega}_1 \wedge d\omega_1 + d{\bar \omega}_2 \wedge d\omega_2 \right)
= d \A
\nonumber\\
\A&=& {i \over 2} \left[
{\bar \omega}_1 \wedge d\omega_1 - \omega_1 \wedge d {\bar \omega}_1
+ {\bar \omega}_2 \wedge d\omega_2 - \omega_2 \wedge d {\bar \omega}_2
\right]\nonumber\\
ds^2&=& d{\bar \omega}_1  d\omega_1 + d{\bar \omega}_2  d\omega_2
\label{mass15}
\eeqar
Using $\omega_1$, $\omega_2$ from (\ref{mass13}), we 
can now separate out the radial coordinate, writing 
\beqar
\omega_1 &=& -i (R_{1i} - i R_{2i} ) x_i = r \bigl[ -i (R_{1i} - i R_{2i} ) \phi_i\bigr]\nonumber\\
\omega_2&=& x_0 + i R_{3i} x_i = r \bigl[ \phi_0 + i R_{3i} \phi_i \bigr]
\label{mass16}
\eeqar
with
$\phi_0 \phi_0 +\phi_i \phi_i = 1$. 
It is straightforward to simplify
$\A$ to get $\A = r^2 \, {\hat \Theta}$,
with
\beqar
{\hat \Theta} &=& R_{3i} ( \phi_i d\phi_0 - \phi_0 d\phi_i )
+ {i \over 2} \left[ (n \cdot \phi) \, d ({\bar n} \cdot \phi) 
- ({\bar n}\cdot \phi)\, d (n \cdot \phi) \right]\nonumber\\
n_i &=& R_{1i }+i R_{2i} = ( - \cos\theta \cos\vf - i \sin\vf, -\cos\theta \sin\vf + i \cos\vf,
\sin\theta )
\label{mass17}
\eeqar
We have chosen a K\"ahler metric for the cone and
we see that $\Omega = d \A $
does have the required structure (\ref{mass11}).
This is basically the Sasakian structure on $S^3$.
Notice that the second term in ${\hat \Theta}$, namely,
\beq
\alpha = {i \over 2} \left[ (n \cdot \phi)\, d ({\bar n} \cdot \phi) 
- ({\bar n}\cdot \phi)\, d (n \cdot \phi) \right]
\label{mass18}
\eeq
defines a local K\"ahler structure for the two-dimensional subspace
transverse to
$R_{3i} \phi_i$. This is only local, since the separation of the third direction
on $S^3$ can only be local. The existence of such a local transverse K\"ahler structure
is a feature of Sasakian manifolds.

The vectors $n_i$, ${\bar n}_i$ define the choice of complex combinations
on the cone. ($R_{3i}$ is not independent, it is proportional to
$({\vec n} \times {\vec{\bar n}})_i$.) A particular K\"ahler structure on the cone
corresponds to a particular choice of these vectors, each of them leading to a
particular Sasakian structure for $S^3$. We have an $S^2$ worth of 
such choices, so the total space we are considering is the 
bundle of Sasakians over $S^3$. In this sense, it is the natural equivalent
in three dimensions of the twistor space in four dimensions\footnote{In Ref.\cite{sasakians}, Boyer and Galicki study a particular
version of what they name as the twistor space
for Sasakian manifolds. They also mention that there could be another object 
which deserves the name of twistor space for Sasakians. The latter one is the trivial
$S^2$ bundle  for a Sasakian manifold inheriting the structure from 
the K\"ahler structure on the cone. It is this latter definition which applies to our case here.}.
Notice that, since $R_{ki}$ is an orthogonal
matrix, 
\beq
n_i n_i = {\bar n}_i {\bar n}_i = 0, \hskip .2in
n_i {\bar n}_i = 2
\label{mass19}
\eeq
In other words, $n_i$, ${\bar n}_i$ are complex null vectors, normalized by the second
relation in (\ref{mass19}). Thus we may think of the bundle of Sasakians
of $S^3$ as the bundle of complex null rays. In this case, the bundle is still trivial
just as it was in ${\mathbb R}^4$.

To proceed further, we introduce stereographic coordinates
$y_i$  for $S^3$ by
\beq
\phi_0 = {y^2 - R^2 \over y^2 + R^2}, \hskip .2in
\phi_i = {2 y_i R \over y^2 + R^2}
\label{mass20}
\eeq
We also introduce the notation
$z = n\cdot y$, $\bz = {\bar n}\cdot y$, $R_{3i} y_i = v$.
The K\"ahler potential for the transverse local K\"ahler one-form $\alpha$ in (\ref{mass18}) is
\beq
K_T = (n\cdot \phi)\,  ({\bar n}\cdot \phi ) = {4 R^2 \bz z \over (\bz z + v^2 + R^2)^2}
\label{mass21}
\eeq
If we take the large $R$ limit, which corresponds to blowing up the $S^3$ to get
${\mathbb R}^3$, we find
\beqar
{\hat \Theta} &\approx& {2 d v \over R} + {2 i \over R^2} ( z d\bz - \bz dz )
\nonumber\\
K_T&\approx& 4 {\bz z \over R^2}
\label{mass22}
\eeqar
This will be useful in applying our results to gauge fields in ${\mathbb R}^3$.

We can now write down the WZW action for the transverse space with the coordinates
$z, \, \bz$, with the (local) K\"ahler metric $(\del \bdel K_T)\,dz d\bz$.
The factors involving $(\del \bdel K_T)$ will drop out of $S_{\rm WZW}$ because of its
 conformal
invariance. The action is thus the integral of a differential two-form
given as
\beqar
S_{\rm WZW} (H) &=& -{i \over 4 \pi} \int_\M \Tr ( \del H \wedge \bdel H^{-1} )
+ {i \over 12 \pi} \int_{\M^3} \Tr( H^{-1} dH \wedge H^{-1} d H \wedge H^{-1} dH )
\nonumber\\
&=& -{i \over 4 \pi} \biggl[
\int_\M \Tr ( \del H \wedge \bdel H^{-1} )\nonumber\\
&&\hskip .05in - \int_{\M^3} \Tr\bigl[ H^{-1} dH \wedge \left(H^{-1} \del H \wedge H^{-1} \bdel H 
- H^{-1} \bdel H \wedge H^{-1} \del H \right)\bigr] \biggr]
\label{mass23}
\eeqar
The mass term of interest can now be written down as $\Gamma = m^2  S_m$, with
\cite{nair-moving}
\beqar
S_m &=& - \int d\mu(S^2)\, {\hat \Theta}\, S_{\rm WZW}(H)\nonumber\\
&=& {i\over 4\pi} \int d\mu(S^2)\, {\hat\Theta}\wedge
\biggl[ \Tr (\del H \wedge \bdel H^{-1})  \nonumber\\
&&\hskip .3in
- \Tr \left[H^{-1} dH \wedge \left(H^{-1} \del H \wedge H^{-1} \bdel H 
- H^{-1} \bdel H \wedge H^{-1} \del H \right)\right] \biggr]
\label{mass24}
\eeqar
The same expression also applies to the large $R$ (or ${\mathbb R}^3$) limit,
where the one-form ${\hat \Theta}$ and the potential $K_T$ simplify as given
 in (\ref{mass22}).
 
The expression (\ref{mass24}) may still seem rather cryptic, but it is straightforward to
work out the expression as a series in terms of the gauge potentials, after Fourier transforming to momentum space. The first two terms
are \cite{alexanian-nair2}
\beqar
S_m &=&{1\over 2}  \int {d^3k \over (2\pi )^3} 
 A_i^a(-k) A_j^a(k)\biggl(\delta_{ij}-{k_ik_j\over
{\vec k}^2}\biggr) \nonumber\\
&&+ \int {d^3k \over (2\pi)^3} {d^3q\over (2\pi)^3} A^a_i (k) A^b_j(q) A^c_k (-k -q) \,f^{abc}
V_{ijk} (k, q, -k-q)\nonumber\\
V_{ijk}(k,q,-(k+q) ) &=& {i \over 6}
\biggl[{1\over k^2q^2-(q\cdot k)^2}\biggr]
\biggl[\biggl\{{q\cdot k\over k^2}-
{q\cdot (q+k)\over (q+k)^2}\biggr\}k_ik_jk_k \nonumber\\
&&\hskip .2in +{k\cdot(q+k)\over (q+k)^2}(q_iq_jk_k+ q_kq_ik_j+q_jq_kk_i)
-(q\leftrightarrow k)\biggr]
\label{mass24a}
\eeqar

\section{Properties of the 3d mass term}

Our arguments in arriving at (\ref{mass24}) show that it has a deep and interesting
mathematical structure and that it is the most natural generalization to three dimensions
of the results in
two and four dimensions.
But we can again ask the crucial question of whether it has anything to do with physics.
Indeed that is the case, the motivation from physics is what led to this
mass term, for ${\mathbb R}^3$, many years ago, although the Sasakian
structure was not clear at that time \cite{nair-moving}.
The general expectation is that in nonabelian gauge theories a mass gap will be dynamically
generated, so potentially, one can get a term like
(\ref{mass24}) in the effective action for such theories. This will be a highly 
nonperturbative result. One can attempt to demonstrate this, and calculate the coefficient
$m^2$, via a gap equation approach where we add and subtract the same term
to the standard Yang-Mills action,
\beqar
S &=& S_{\rm YM} + m^2 S_m - \Delta S_m = {\tilde S} - \Delta S_m\nonumber\\
{\tilde S}&=&S_{\rm YM} + m^2 S_m
\label{mass26}
\eeqar
The idea is to consider $m^2$ as the exact value of the mass generated
by interactions
while 
$\Delta$ is taken to have a loop expansion of the form
$\Delta=\Delta^{(1)}+\Delta^{(2)}+\ldots$. Calculations can be done in a
loop expansion, with the action ${\tilde S}$ used to construct the 
propagators and vertices at the tree level, but $\Delta$ starts at the one-loop level.
Since $m^2$ is taken to be the exact dynamically generated mass,
the pole of the propagator must remain at
$k_0^2-{\vec k}^2= m^2$
as loop corrections are added. This requires choosing $\Delta^{(1)}$
to cancel the one-loop shift of the pole, $\Delta^{(2)}$ to cancel
the two-loop shift of the pole, etc., as is
usually done for mass renormalization. 
After this is done, the $\Delta$ so obtained must still equal $m^2$,
since the theory is defined by just the YM action.
Thus we must impose
the condition
\beq
\Delta=\Delta^{(1)}+\Delta^{(2)}+\ldots= m^2
\label{mass27}
\eeq
This statement of equating the corrections to the mass
term to $m^2$ is the gap equation
which determines $m^2$ \cite{alexanian-nair2}. (This strategy can be continued to arbitrary orders of calculation \cite{{JP-magmass},{alexanian-nair2}}.)
Notice that the approach is completely gauge-invariant.
The calculation of $m^2$ along these lines was carried out in Ref.\cite{alexanian-nair2}
and gave the value $m \approx 1.19 \, (e^2 c_A /2\pi)$, where $e$ is the gauge coupling and
$c_A$ is the quadratic Casimir 
value for the adjoint representation of the gauge group.
(For other related approaches to the magnetic screening mass, see
Refs.\cite{magmass}, \cite{JP-magmass}.)

The gap equation can be viewed as the result of the summation of an infinite number of Feynman diagrams, a particular sequence being chosen by the form of the mass term. A very different approach
is to use the Schr\"odinger equation in the Hamiltonian formulation of
the theory and to solve it for the ground state
wave functional in a low energy approximation. Such an approach, which has been 
developed in a series
of articles\cite{KKN}, leads to a prediction for the string tension (which is in good agreement with lattice
simulations \cite{teper}) and a value for $m$ as $e^2 c_A/2\pi$. This is close to the value obtained from
the gap equation analysis. 
Yet another independent validation of the result comes from using the same
Hamiltonian approach to calculate the Casimir energy
for two parallel plates \cite{KN2}.  One can then obtain a direct and independent
numerical estimate of the value of the mass by
a lattice simulation of the parallel plate geometry for the Yang-Mills theory.
Such a simulation yields the same value of
$e^2 c_A/2\pi$ to within a fraction of a percent \cite{chernodub}.

There are two other observations regarding this mass term which might be interesting.
The first is about the one-loop correction
generated
by the mass term which determines the gap equation.
Let us denote, in any gauge theory,
the correction to the two-point function as
$\int A_i^a(-k) \Pi_{ij} (k) A^a_j(k)$. Then one can analyze some of the excitations
which can occur in intermediate states of the one-loop graph, 
corresponding to a unitarity cut of the
diagram, by the singularity structure of $\Pi_{ij}(k)$.
As an example, if we use a mass term 
\beq
S_m =  \int d^3x \Tr \left[ F_i \left( {1\over D^2} \right) F_i\right], \hskip .2in
F_i = {1\over 2} \epsilon_{ijk} F_{jk},
\label{mass28}
\eeq
one can show that there are singularities at $k ^2 = 0$ in $\Pi_{ij}(k)$ indicating that there are still zero-mass excitations present in the spectrum \cite{JP-magmass}.
 One way to understand this is to note that if we 
write (\ref{mass28}) in terms of local monomials with an auxiliary field, we get
\beq
S_m = - \int d^3x \Tr \left[ {1\over 2} \phi_i\, (-D^2) \phi_i + \phi_i F_i \right]
\label{mass29}
\eeq
This would give propagating massless solutions corresponding to
$\,\Box\, \phi_i = 0$ in the absence of $A_i$. In contrast to this, 
$\Pi_{ij}(k)$ resulting from the mass term (\ref{mass24}) or (\ref{mass24a}) has no zero-mass
threshold singularities, at least to the one-loop order the calculations have been carried out.
One way to understand this may be to note that the mass term rewritten in terms of local
monomials
with auxiliary fields is
\beqar
S_m &=& \int d\mu(S^2) \biggl[
\int dx^T S_{\rm WZW} (G) \nonumber\\
&&+ {1\over \pi} \int d^3x\, \Tr \left( G^{-1} \bdel G \, A
- \bA\,  \del G G^{-1} + A\, G^{-1} \bA \,G - A \bA \right)
\biggr]
\label{mass30}
\eeqar
The equations of motion for the group element $G$ leads to the solution
\beq
G = M^{\dagger -1} {\tilde V} (z)  \,V(\bz ) M^{-1}
\label{mass31}
\eeq
Since the matrices $M$, $M^\dagger$ are only defined
up to the ambiguity $M \rightarrow M V^{-1}(\bz )$,
$M^\dagger \rightarrow {\tilde V}( z) M^\dagger$ by the equations
$A = -\del M\,M^{-1}$, $\bA = M^{\dagger -1} \bdel M^\dagger$,
we can absorb ${\tilde V} (z)\,V(\bz )$ in (\ref{mass31}) into the definition of
$M$, $M^\dagger$. Thus there are no independent solutions, or independent degrees of freedom,
for the
auxiliary field.

Our second observation is about the use of this mass term in the context of the 
quark-gluon plasma.
The hard thermal loops generate a screening mass (\ref{mass5}) or (\ref{mass8})
for the chromoelectric forces,
the mass term (\ref{mass24}) can describe the magnetic screening or the magnetic mass
of the plasma. In carrying out calculations at finite temperature, one can see that, even after
taking account of the hard thermal loops and the corresponding chormoelectric screening effects,
there are still infrared divergences left over in the nonabelian theory.
These are cured by the screening mass for the chromomagnetic interactions, 
so the dynamical generation of such a mass term is an important feature.
At very high temperatures, the 3+1 dimensional theory can be approximated by
the same theory in three Euclidean dimensions, i.e., there is a dimensional reduction,
the coupling of the 3d theory being
$e^2 = g^2 T$, where $g$ is the 4d coupling. 
The dynamically generated mass of the 3d theory can thus be interpreted as
the magnetic screening mass of the high temperature limit of the
4d theory \cite{GPY}.
For this idea to be implemented in the full four-dimensional theory, we again
need a 4d-Lorentz invariant form of the mass term.
It is indeed possible to construct such an invariant mass term \cite{nair-moving}.
The result is exactly of the form given in
(\ref{mass8}) with one change. Instead of the components of $A_\mu$
given (\ref{mass9}), we must use
\beq
A_\pi = {\half} \pi (e\cdot A) {\bar \xi} , \hskip .2in
A_\xi = {\half} \xi (e \cdot A) {\bar \pi}
\label{mass32}
\eeq
Notice that in each of these combinations there is mixing of the spinors $\pi$, $\xi$, unlike
the situation in ({\ref{mass9}).

\section*{Acknowledgements}

I thank A.P. Balachandran and Denjoe O'Connor for discussions
on the Sasakian structure.
This research was supported in part by the U.S.\ National Science
Foundation grant PHY-1820721
and by PSC-CUNY awards.



\begin{thebibliography}{99}
\bibitem{JP-soliton} R. Jackiw and S-Y. Pi, \PRL~{\bf64}, 2969 (1990).

\bibitem{JN1} R. Jackiw and V.P. Nair, \PR~{\bf D43}, 1933 (1991); 
\PRL~{\bf 73}, 2007 (1994); \PL~{\bf B480}, 237 (2000); \PL~{\bf B551}, 166 (2003).

\bibitem{JN-clebsch} R. Jackiw, V.P. Nair and S-Y. Pi, \PR~{\bf D62}, 085018 (2000).

\bibitem{hydro} B. Bistrovic {\it et al}, \PR~{\bf D67}, 025013 (2003);
R. Jackiw, V.P. Nair, S-Y. Pi and A.P. Polychronakos, 
{\it J. Phys. A: Math. Gen.} {\bf 37}, R327 (2004).

\bibitem{nair-htl} R. Efraty and
V.P. Nair, \PRL~ {\bf 68}, 2891 (1992);
\PR~ {\bf D47}, 5601 (1993).

\bibitem{JN-htl2} R. Jackiw and V.P. Nair, \PR~ {\bf D48}, 4991 (1993). 

\bibitem{JP-magmass} R. Jackiw and So-Young Pi, \PL~{\bf B368}, 131 (1996);
{\it ibid.}~{\bf B403}, 297 (1997).
 
\bibitem{witten} E. Witten, ~\CMP, {\bf 92}, 455 (1984); 
S.P. Novikov, {\it Usp. Mat. Nauk} {\bf 37}, 3 (1982).

\bibitem{PW} A.M. Polyakov and P.B. Wiegmann, \PL~ {\bf B141}, 223 (1984).

\bibitem{schwinger} J.~S.~Schwinger, \PR~{\bf 128}, 2425 (1962).

\bibitem{twistor} R. Penrose,
\JMP\,  {\bf8}, 345 (1967). The subject is by now very well developed,
see R. Penrose and M.A.H. MacCallum, {\it Phys. Rep.}
{\bf 6}, 241 (1972); R. Penrose and W. Rindler,
\textit{Spinors and Spacetime}, 2 volumes (Cambridge University
Press, 1984 \& 1987);
S.A. Hugget and K.P. Tod, \textit{An Introduction to Twistor Theory}
(Cambridge University Press, 1993).

\bibitem{htl} R. Pisarski, {\it Physica} {\bf A158}, 246 (1989); 
\PRL~ {\bf 63}, 1129 (1989);
E. Braaten and R. Pisarski, \PR~ {\bf D42}, 2156 (1990);
\NP~ {\bf B337}, 569 (1990); {\it
ibid.} {\bf B339}, 310 (1990); \PR~ {\bf
D45}, 1827 (1992);
J. Frenkel and J.C. Taylor, \NP~
{\bf B334}, 199 (1990); J.C. Taylor and S.M.H. Wong, 
\NP~ {\bf B346}, 115 (1990).

\bibitem{alexanian-nair} G. Alexanian and V.P. Nair, 
\PL~ {\bf B390}, 370 (1997).

\bibitem{nair-moving} V.P. Nair,  \PL~ {\bf B352}, 117 (1995).

\bibitem{sasakians} S. Sasaki, {\it Tohoku Math. J.} {\bf 2}, 459 (1960);
For further developments and reviews, see C. Boyer and K. Galicki,
{\it Surveys Diff. Geom.} {\bf 7}, 123 (1999) [arXiv:hep-th/9810250].

\bibitem{alexanian-nair2} G. Alexanian and V.P. Nair,  \PL~ {\bf B352},  435 (1995).

\bibitem{magmass} 
J.M. Cornwall, \PR~ {\bf D10}, 500 (1974); 
{\it ibid.} {\bf D26}, 1453 (1982); {\it ibid.} {\bf D57}, 3694 (1998);
W. Buchmuller and O. Philipsen, \NP~{\bf B443} (1995) 47;
O. Philipsen, in {\it TFT-98: Thermal Field Theories and
their Applications}, U. Heinz (ed.), hep-ph/9811469;
J. M. Cornwall and B. Yan, \PR~{\bf D53}, 4638 (1996);
J.M. Cornwall, \PR~{\bf D76}, 025012 (2007).


\bibitem{KKN} D. Karabali and V.P. Nair,
\NP~ {\bf B464}, 135 (1996);
\PL~ {\bf B379}, 141 (1996);
D. Karabali, Chanju Kim and V.P. Nair, \NP~ {\bf B524}, 661 (1998);
\PL~ {\bf B434}, 103 (1998); D. Karabali, V.P. Nair and A. Yelnikov, \NP~ {\bf B824}, 387 (2010);
For a short review, see V.P. Nair, arXiv:0910.3252,
{\it Proceedings of Science, POS(QCD-TNT09) 030}, 
\verb+http://pos.sissa.it//archive/conferences/087/030/QCD-TNT09_030.pdf+.

\bibitem{teper} M. Teper, 
\PR ~{\bf D59}, 014512 (1999) and references therein;
B. Lucini and M. Teper, \PR ~{\bf D66}, 097502 (2002);
 B.~Bringoltz and M.~Teper,
\PL~ {\bf B645}, 383 (2007).

\bibitem{KN2} D. Karabali and V.P. Nair, \PR~{\bf D98}, 105009 (2018).

\bibitem {chernodub} M.N. Chernodub, V.A. Goy, A.V. Molochkov and
H.H. Nguyen, \PRL~{\bf 121}, 191601 (2018).

\bibitem{GPY} See, for example,
D. Gross, R. Pisarski and L. Yaffe, \RMP~ {\bf 53}, 43 (1981) and references therein.


\end{thebibliography}
\end{document}